\documentstyle[12pt,psfig]{article}
\begin{document}
\author{Santi Prestipino \\
Istituto Nazionale per la Fisica della Materia (INFM) and \\
Universit\`a degli Studi di Messina, Dipartimento di Fisica, \\
Contrada Papardo, 98166 Messina, Italy; \\
e-mail: {\tt Santi.Prestipino@unime.it}}
\title{A probabilistic model \\
for the equilibration of an ideal gas}
\maketitle
\begin{abstract}
~~I present a generalization of the Ehrenfest urn model that is
aimed at simulating the approach to equilibrium in a dilute gas.
The present model differs from the original one in two respects:
1) the two boxes have different volumes and are divided into
identical cells with either multiple or single occupancy;
2) particles, which carry also a velocity vector, are subjected to
random, but elastic, collisions, both mutual and against the container
walls.
I show, both analytically and numerically, that the number and energy
of particles in a given urn evolve eventually to an equilibrium
probability density
$W$ which, depending on cell occupancy, is binomial or hypergeometric
in the particle number and beta-like in the energy.
Moreover, the Boltzmann entropy $\ln W$ takes precisely the same form
as the thermodynamic entropy of an ideal gas.
This exercise can be useful for pedagogical purposes in that it
provides, although in an extremely simplified case, a probabilistic
justification for the maximum-entropy principle.

\vspace{4mm}

\vspace{4mm}
\end{abstract}
\newpage
\section{Introduction}
The modern intuition of the emergence of the Second Law of thermodynamics
from mechanics is mainly grounded upon the behaviour of stochastic urn
models, where ``particles'' are subjected to a {\em probabilistic dynamics}
that eventually generates a sort of thermodynamic equilibrium~\cite{Ehrenfest}.
Obviously, this stochastic (Markovian) dynamics is only a caricature of
the ``real'' (Newtonian) dynamics; it is much like an effective dynamics
which emerges after averaging over many instances of the complicated
short-time motion.

In the classical Ehrenfest model, $N$ numbered balls are distributed
into two urns; at each step of the process, a ball is extracted at
random and moved from the urn where it resides to the other.
Eventually, the number of balls in each urn fluctuates around $N/2$,
with relative deviations from the mean becoming negligible in the
large-$N$ limit.
This stochastic process is taken to represent the attainment of
particle-number equilibrium in a dilute gas diffusing between two
communicating vessels of equal volume.

In order to improve the Ehrenfest model so as to make it more realistic,
I consider a generalization where the balls/particles are endowed with
both a discrete position and a continuous velocity.
To be more specific, we are given two boxes, 1 and 2, and $N$ particles
in the boxes.
Box 1 (2) is divided into $V_1$ ($V_2$) identical cells, with $V=V_1+V_2$
the total cell number.
The occupancy $c_{\alpha}$ of the $\alpha$-th cell can be either multiple
($c_{\alpha}=0,1,2,\ldots$) or single ($c_{\alpha}=0,1$), with both
possibilities being considered.
The velocity of the $a$-th particle is $v_a$, a three-dimensional vector
with components $v_{ak}$ ($k=1,2,3$).

To make things simple, positions and velocities are updated independently
and by turns, in such a way that the two dynamics of free motion and
collisions will proceed in parallel though staying separate.
Hence, the stationary state of each set of variables can be analysed
on its own.
Along this route, one arrives at a probabilistic foundation of the
expression of the ideal-gas entropy in thermodynamics and, concurrently,
at a justification (in this case only) of the maximum-entropy principle.

\section{Position updates}
Let us first suppose that each cell can host whatever number of particles.
A position update consists of i) choosing at random one particle, $a_r$,
and one cell, $\alpha_r$; and ii) moving $a_r$ from its original cell
into $\alpha_r$.
In terms of the variable $n$, which counts how many particles are hosted in
box 1, this defines a stochastic process of the Markov type, with
transition probabilities:
\begin{equation}
T(n+1\leftarrow n)=\frac{(N-n)V_1}{NV}\,;\,\,\,\,\,\,
T(n-1\leftarrow n)=\frac{nV_2}{NV}\,.
\label{eq01}
\end{equation}
The ensuing master equation admits the {\em binomial} distribution
\begin{equation}
W(n)={N\choose n}\left( \frac{V_1}{V}\right) ^n
\left( \frac{V_2}{V}\right) ^{N-n}\,
\label{eq02}
\end{equation}
as unique stationary distribution~\cite{Prestipino}.
Since the Markov chain is ergodic ({\it i.e.}, there is a path connecting
every (macro)state $n$ to every other $n^{\prime}$), any initial
distribution $P(n;0)$ will converge, in the long run, to $W(n)$.
In particular, the average $n$ goes eventually into $NV_1/V$, with
relative deviations from the mean of ${\cal O}(N^{-1/2})$~\cite{Prestipino}.
The multiplicity of macrostate $n$, {\it i.e.}, the number of complexions
(microstates) of $N$ numbered particles in the boxes, such that box 1
contains $n$ particles, is just $W(n)\times V^N$.
Assuming $n,N-n={\cal O}(N)\gg 1$, the equilibrium entropy $S(n)$,
defined as the logarithm of the multiplicity, is additive over the
boxes and extensive with $N$:
\begin{equation}
S(n)\sim -n\ln\frac{n}{V_1}-(N-n)\ln\frac{N-n}{V_2}\,,
\label{eq03}
\end{equation}
being maximum for $n=NV_1/V$.
In Eq\,(\ref{eq03}), we recognize the volume contribution to the
ideal-gas entropy.

While all of the above is rather standard, a novel result is obtained
when each cell in the boxes can contain at most one particle.
Now, at each step in the process, the selected particle $a_r$ is moved
into a cell $\alpha_r$ that is chosen at random among the vacant sites.
The transition probabilities read (with $V_1,V_2\geq N$):
\begin{equation}
T(n+1\leftarrow n)=\frac{(N-n)(V_1-n)}{N(V-N)}\,;\,\,\,\,\,\,
T(n-1\leftarrow n)=\frac{n(V_2-N+n)}{N(V-N)}\,,
\label{eq04}
\end{equation}
yielding a {\em hypergeometric} stationary distribution for
$n$~\cite{Prestipino}:
\begin{equation}
W(n)={V\choose N}^{-1}{V_1\choose n}{V_2\choose N-n}\,.
\label{eq05}
\end{equation}
As in the previous case, the average $n$ converges eventually to $NV_1/V$,
with relative deviations of ${\cal O}(N^{-1/2})$.
The multiplicity of state $n$, {\it i.e.}, the number of ways $N$
indistinguishable particles can be arranged in the boxes, in such a way
that $n$ particles are placed in box 1, is equal to $W(n)\times{V\choose N}$.
Assuming $n,N-n,V_1-n,V_2-(N-n)={\cal O}(N)\gg 1$, the equilibrium entropy
becomes:
\begin{eqnarray}
S(n) &\sim& -n\ln\frac{n}{V_1}-(V_1-n)\ln\left( 1-\frac{n}{V_1}\right)
\nonumber \\
&-& (N-n)\ln\frac{N-n}{V_2}-(V_2-N+n)\ln\left( 1-\frac{N-n}{V_2}\right) \,,
\label{eq06}
\end{eqnarray}
being maximum for $n=NV_1/V$.
Equation (\ref{eq06}) is nothing but the thermodynamic entropy of two
ideal {\em lattice} gases that can mutually exchange energy and particles.

\section{Velocity updates}
The collision dynamics of equal-mass particles can be roughly schematized
as a succession of {\em random} binary events which, however, are still
required to obey energy and momentum conservation~\cite{Sauer}.
On the macroscopic side, such collision rules go along with the conservation
of total {\em kinetic} energy and total momentum, thus being appropriate
only to a very dilute (gaseous) system of particles.
If, moreover, we are willing to drop the momentum constraint, provision
should be made also for elastic collisions against the walls of the
(cubic) container.

As far as the mutual collisions are concerned, I assume their outcome
to be as maximally random as possible.
This amounts to update the velocities of the colliding particles as:
\begin{equation}
v_a\rightarrow v_a^{\prime}=v_a+\xi\,\hat{r}\,;\,\,\,\,\,\,
v_b\rightarrow v_b^{\prime}=v_b-\xi\,\hat{r}\,,
\label{eq07}
\end{equation}
where $\xi=(v_b-v_a)\cdot\hat{r}$, and $\hat{r}$ is picked up
at random out of the emisphere of unit-length vectors forming an
acute angle with $v_b-v_a$.
If mutual collisions occur at a rate of $p$, the master equation for
the velocities finally reads:
\begin{equation}
\pi(\{v^{\prime}\};t+1)=\int{\rm d}^{3N}v\,
\tau(\{v^{\prime}\}\leftarrow\{v\})\,\pi(\{v\};t)\,,
\label{eq08}
\end{equation}
with $\tau=(1-p)\tau_1+p\tau_2$ and
\begin{eqnarray}
\tau_1(\{v^{\prime}\}\leftarrow\{v\}) &=& \frac{1}{3N}
\sum_{a=1}^N\sum_{k=1}^3\left[ \delta(v_{ak}^{\prime}+v_{ak})
\prod_{(b,\,l)\neq(a,\,k)}\delta(v_{bl}^{\prime}-v_{bl})\right] \,;
\nonumber \\
\tau_2(\{v^{\prime}\}\leftarrow\{v\}) &=& \frac{2}{N(N-1)}\sum_{a<b}
\left[ \frac{1}{2\pi\left| v_a-v_b\right| }
\delta^3(v_a^{\prime}+v_b^{\prime}-v_a-v_b)
\delta(v_a^{\prime 2}+v_b^{\prime 2}-v_a^2-v_b^2)\right.
\nonumber \\
&\times&\left.
\prod_{c\neq a,\,b}\delta^3(v_c^{\prime}-v_c)\right] \,.
\label{eq09}
\end{eqnarray}

The following properties can be proved~\cite{Prestipino}:

$\bullet$ A stationary solution to Eq.\,(\ref{eq08}) is
$w(\{v\})=F(v_1^2+\ldots+v_N^2)$, for any properly normalized function $F$.

$\bullet$ Upon denoting the one- and two-body velocity distributions at
time $t$ as $f_1(v_1;t)$ and $f_2(v_1,v_2;t)$, the following exact
equation of evolution holds:
\begin{eqnarray}
f_1(v_1;t+1) &=& (1-p)\left\{ \left( 1-\frac{1}{N}\right) f_1(v_1;t)\right.
\nonumber \\
&+& \left. \frac{1}{3N}\left[ f_1(-v_{1x},v_{1y},v_{1z};t)+
f_1(v_{1x},-v_{1y},v_{1z};t)+f_1(v_{1x},v_{1y},-v_{1z};t)\right] \right\}
\nonumber \\
&+& p\left\{ \left( 1-\frac{2}{N}\right) f_1(v_1;t)
+\frac{2}{N}\times\frac{1}{2\pi}\int{\rm d}^3v_2\int{\rm d}^3\Delta
\frac{1}{\left| \Delta\right|}\,%
\delta\left[ \Delta^2-\left( \frac{v_1-v_2}{2}\right) ^2\right] \right.
\nonumber \\
&\times& \left. f_2\left( \frac{v_1+v_2}{2}+\Delta,\frac{v_1+v_2}{2}-
\Delta;t\right) \right\} \,.
\label{eq10}
\end{eqnarray}
For any function $\Phi$, the {\em ansatz}
$f_2^{\rm (eq)}(v_1,v_2)=\Phi(v_1^2+v_2^2)$ gives a stationary solution to
Eq.\,(\ref{eq10}).
However, in case of an isolated system with total energy $U$, the only
admissible solution to Eq.\,(\ref{eq08}) is the microcanonical density
$w(\{v\})\propto \delta(v_1^2+\ldots +v_N^2-U)$~\cite{note}, and the
$\Phi$ function becomes:
\begin{equation}
f_2^{\rm (eq)}(v_1,v_2)=\frac{\Gamma(3N/2)}{\Gamma(3(N-2)/2)}(\pi U)^{-3}
\left( 1-\frac{v_1^2+v_2^2}{U}\right) ^{\frac{3(N-2)}{2}-1}\,,
\label{eq11}
\end{equation}
leading in turn to:
\begin{equation}
f_1^{\rm (eq)}(v_1)=\frac{\Gamma(3N/2)}{\Gamma(3(N-1)/2)}(\pi U)^{-\frac{3}{2}}
\left( 1-\frac{v_1^2}{U}\right) ^{\frac{3(N-1)}{2}-1}\,.
\label{eq12}
\end{equation}
The latter is the finite-$N$ Maxwell-Boltzmann (MB) distribution~\cite{Mello}.
In the $N,U\rightarrow\infty$ limit (with $U/N={\cal O}(1)$), one recovers
from Eq.\,(\ref{eq12}) the more familiar Gaussian form,
$f_1^{\rm (eq)}(v)=(\kappa/\pi)^{3/2}\exp(-\kappa v^2)$, with
$\kappa=3N/(2U)$ (corresponding to an average $v_a^2$ of $U/N$
for all $a$).

$\bullet$ I have carried out a computer simulation of the evolution
encoded in Eq.\,(\ref{eq08}) in order to check whether the stationary
distribution (\ref{eq12}) is also an asymptotic solution, as expected
(at least when $p>0$) from the ergodicity of kernel (\ref{eq09}).
First, I set $N=3$ and $U=0.06$, with $p=0.5$.
Starting at any particular microstate with energy $U$, I collect in a
hystogram the values, at regular time intervals, of the three components
of particle-1 velocity (see Fig.\,1 left).
Indeed, this hystogram has, in the long run, the finite-$N$ MB form.
This is indirect evidence that the simulation trajectory samples
uniformly, at least effectively if not literally, the $3N$-dimensional
hypersurface of energy $U$.

Afterwards, I take $N=1000$ and $U=20$, and follow the evolution of the
same hystogram as above, now starting from velocity values extracted at
random from {\it e.g.} a uniform one-particle distribution of variance
$U/(3N)$.
The long-run distribution of velocity no.\,1 compares well with a
Gaussian (Fig.\,1 right), that is with the large-$N$ form of the MB
distribution.
In fact, also the instantaneous velocities of all particles are
asymptotically distributed, for large $N$, according to the same
Gaussian (see Fig.\,2).
This indicates that: 1) the vast majority of points in the energy
hypersurface is made of ``typical'' states, {\it i.e.},
microstates that look more or less similar as far as low-order
distributions like $f_1$ are concerned; and 2) the
microstate at which the evolution was started is, indeed,
untypical~\cite{Goldstein}.

$\bullet$ For a given number $n$ of particles in box 1, the equilibrium
probability density $W_n(u)$ of their total energy $u$ can be calculated
exactly for $w(\{v\})\propto\delta(v_1^2+\ldots +v_N^2-U)$:
\begin{equation}
W_n(u)=\frac{\Gamma(3N/2)}{\Gamma(3n/2)\,\Gamma(3(N-n)/2)}\,
U^{-\left( \frac{3N}{2}-1\right) }
u^{\frac{3n}{2}-1}(U-u)^{\frac{3(N-n)}{2}-1}\,,
\label{eq13}
\end{equation}
that is, variable $u/U$ is {\em beta}-distributed with an average of $n/N$.
Of all $n$-velocity microstates, the fraction of those states whose energy
lies between $u$ and $u+\Delta u$ is $W_n(u)\Delta u$ ($\Delta u\ll u$).
In particular, the Boltzmann entropy associated with Eq.\,(\ref{eq13}) is,
for $n,N-n={\cal O}(N)\gg 1$:
\begin{equation}
\ln W_n(u)\sim -\frac{3n}{2}\ln\frac{n}{u}-\frac{3(N-n)}{2}\ln\frac{N-n}{U-u}\,,
\label{eq14}
\end{equation}
which, when including also the configurational term (\ref{eq03}) or
(\ref{eq06}), gives back the expression of the entropy of the
(monoatomic) ideal gas.

In conclusion, I have introduced a stochastic process of the Ehrenfest
type which allows one to found microscopically the expression of the
thermodynamic entropy of an ideal gas, without relying on the hypothesis
of equal {\it a priori} probability of all microstates.
Rather, the validity of the latter, at least in an effective sense,
follows as an outcome from the stochastic dynamics itself.
In thermodynamics, the Second Law requires the maximization of the
total entropy $S$ under the given constraints (here, the total number
of particles $N$ and the total energy $U$ of two ideal gases in
grand-canonical contact) in order to find the equilibrium state of
an overall isolated system.
In the present model, this very same prescription emerges naturally,
when defining entropy {\it \`a la} Boltzmann, as the condition upon
which the partition of $N$ and $U$ between the gases be, in the
long-time regime, the most probable.
\newpage
%
%

\newpage
%
%
\begin{center}
\large
FIGURE CAPTIONS
\normalsize
\end{center}
\begin{description}
\item[{\bf Fig.\,1 :}]
Numerical simulation of Eq.\,(\ref{eq08}).
Hystogram of velocity values for particle 1 ($\triangle$, $\Box$, and
$\bigcirc$ correspond to the $x,y$, and $z$ component, respectively).
Two distinct values of $N$ are compared, {\it i.e.}, 3 (left) and
1000 (right), while $U/N=0.02$ in both cases.
After rejecting a total of $10^4$ collisions per particle (CPP), as
many as ${\cal N}_{\rm eq}$ CPP are produced
(${\cal N}_{\rm eq}=10^7$ for $N=3$ and ${\cal N}_{\rm eq}=10^6$ for
$N=1000$).
The $p$ value was 0.5, held fixed during the simulation.
Data (in form of frequencies of occurrence) are grouped in bins of width
$\delta v=2\sqrt{U/N}/31$ (after equilibration, the hystogram is updated
every 10 CPP).
The full curve is the theoretical, finite-$N$ MB distribution per
velocity component, which, for $N=3$, is appreciably different
from the infinite-$N$ limit ({\it i.e.}, the Gaussian
$\sqrt{\kappa/\pi}\exp(-\kappa v^2)$, with $\kappa=3N/(2U)$ -- broken
curve in the left panel, full curve in the right panel).

\item[{\bf Fig.\,2 :}]
Numerical simulation of Eq.\,(\ref{eq08}).
Top: Particle velocities at the end of the simulation run for $N=1000$
and $U=20$ (same symbols and notation as in Fig.\,1).
The distribution of all-particle velocities at a given time strongly
resembles the same Gaussian as in Fig.\,1 (full curve).
Bottom: Difference between the above hystogram and this Gaussian law.
\end{description}
\newpage
%
%
\begin{figure}
\begin{center}
\setlength{\unitlength}{1cm}
\begin{picture}(18,10)(0,0)
\put(-1.5,0){\psfig{file=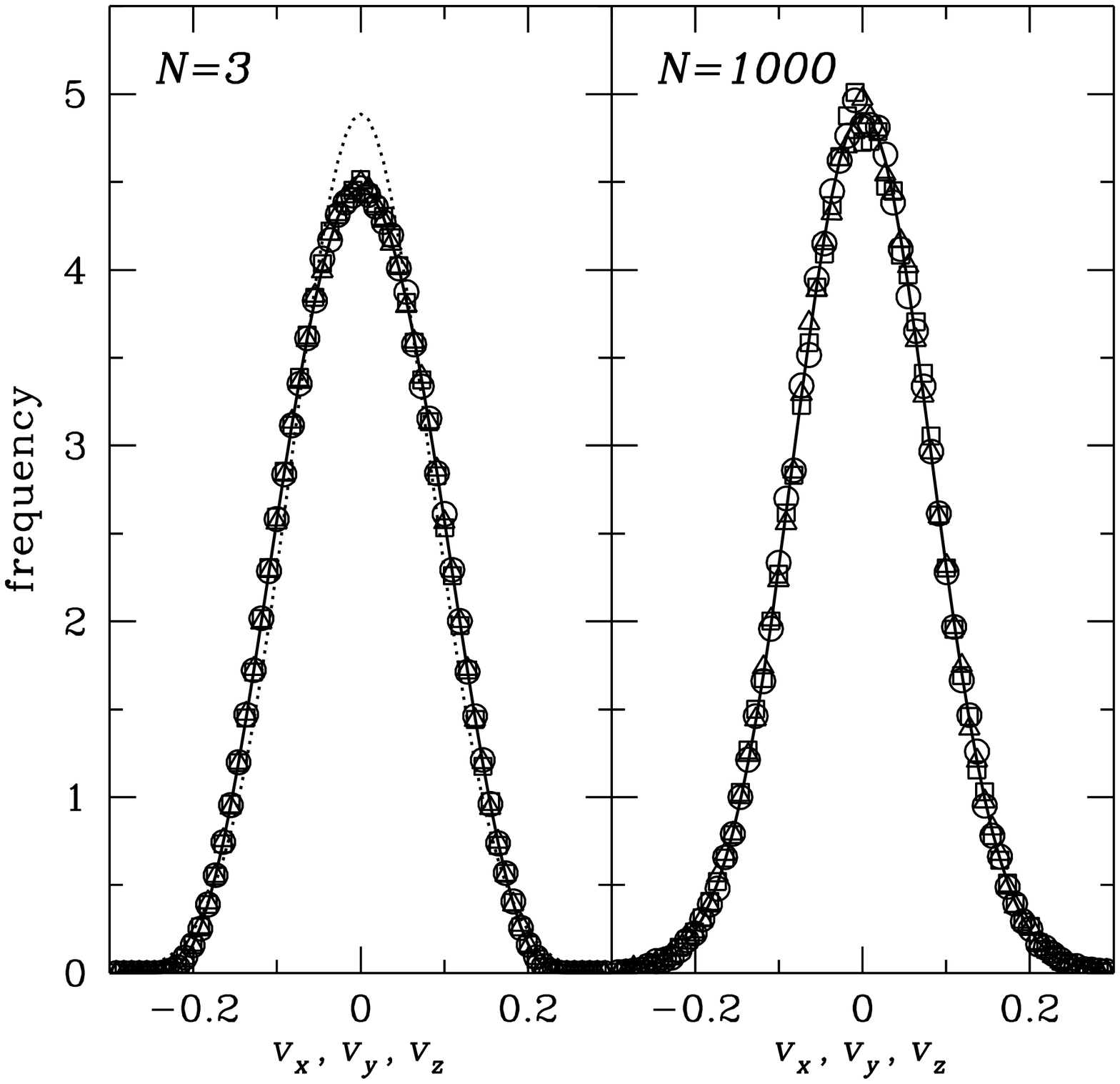,width=18cm,bbllx=0cm}}
\end{picture}
\caption[1]{
}
\end{center}
\end{figure}
%
%
\begin{figure}
\begin{center}
\setlength{\unitlength}{1cm}
\begin{picture}(18,10)(0,0)
\put(-1.5,0){\psfig{file=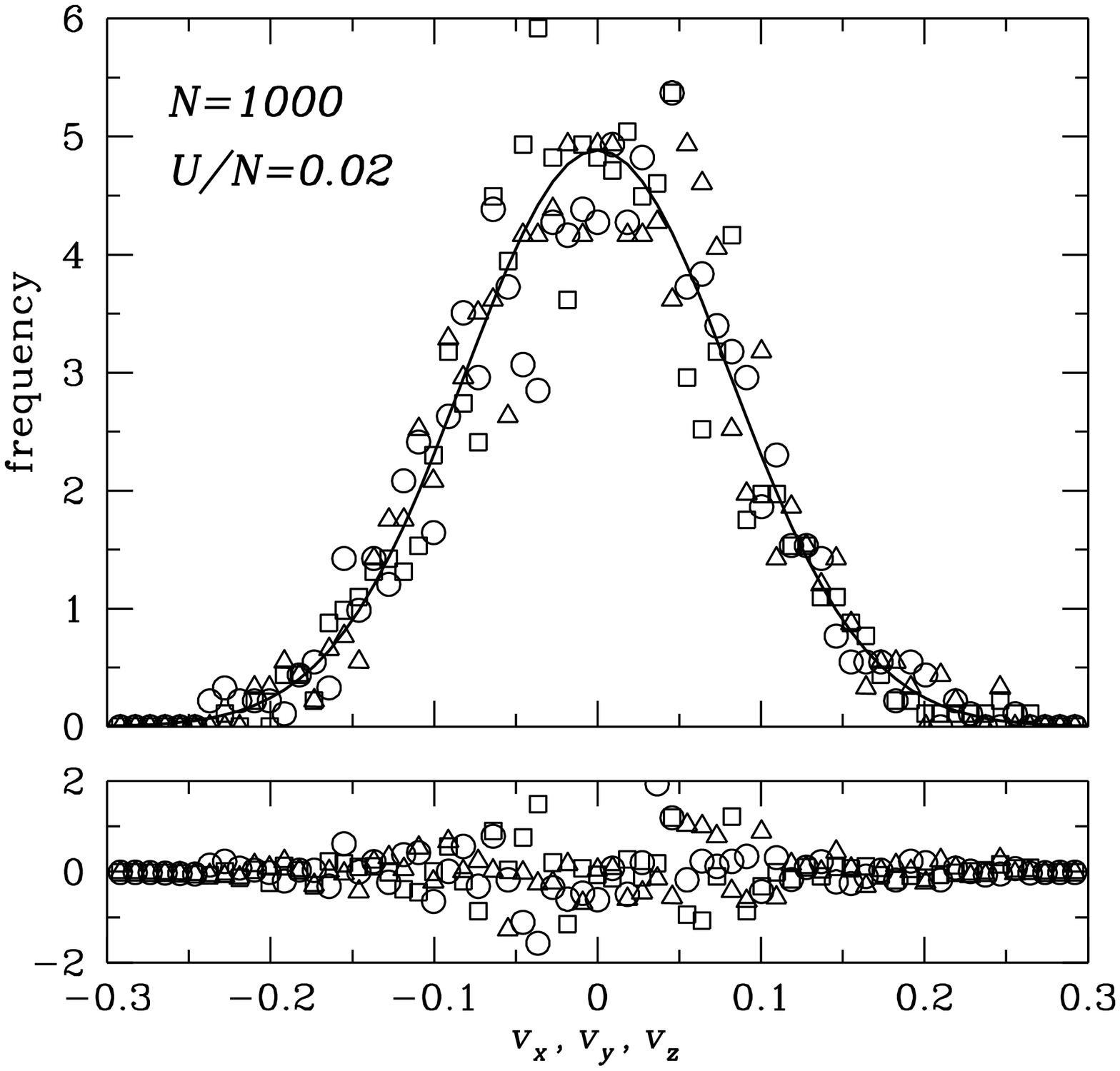,width=18cm,bbllx=0cm}}
\end{picture}
\caption[2]{
}
\end{center}
\end{figure}

\begin{thebibliography}{99}
\bibitem{Ehrenfest}  The forerunner of all urn models is found
in P. Ehrenfest and T. Ehrenfest, {\it Physik. Z.} {\bf 8}, 311 (1907).
For more recent work on urn models, see Y.-M. Kao and P.-G. Luan,
{\it Phys. Rev. E} {\bf 67}, 031101 (2003), and references therein.

\bibitem{Prestipino}  S. Prestipino, {\it Phys. Rev. E} (2003), submitted.

\bibitem{Sauer}  For related work, see {\it e.g.} G. Sauer,
{\it Am. J. Phys.} {\bf 49}, 13 (1981); M. Eger and M. Kress,
{\it Am. J. Phys.} {\bf 50}, 120 (1982); D. Blackwell and
R. D. Mauldin, {\it Lett. Math. Phys.} {\bf 10}, 149 (1985).

\bibitem{note}  $U$ is meant to express the value of the total kinetic
energy in units of $m/2$, $m$ being the particle mass.

\bibitem{Mello}  P. A. Mello and T. A. Brody, {\it Am. J. Phys.} {\bf 40},
1239 (1972).

\bibitem{Goldstein}  See these points carefully discussed in the
contributing paper by S. Goldstein to {\em Chance in Physics:
Foundations and Perspectives} (J. Bricmont {\em et al.} eds., Springer, 2001).
\end{thebibliography}
\end{document}